\documentclass[12pt,preprint]{aastex}

\begin{document}

\title{The Shapes of Galaxies in the Sloan Digital Sky Survey} 
\author{S. M. Khairul Alam \& Barbara S. Ryden
}

\affil{Department of Astronomy, The Ohio State University}
\affil{140 W. 18th Avenue, Columbus, OH 43210\\
     alam, ryden@astronomy.ohio-state.edu
}

\begin{abstract}
We estimate the distribution of apparent axis ratios $q$ for
galaxies in the Sloan Digital Sky Survey (SDSS) Early Data
Release. We divide the galaxies by profile type (de Vaucouleurs
versus exponential) as well as by color ($u^* - r^* \leq 2.22$
versus $u^* - r^* > 2.22$). The axis ratios found by fitting
models to the surface photometry are generally smaller
than those found by taking the second moments of the
surface brightness distribution. Using the axis ratios
found from fitting models, we find that galaxies with
de Vaucouleurs profiles have axis ratio distributions which
are inconsistent, at the 99\% confidence level, with their
being a population of randomly oriented oblate spheroids. 
Red de Vaucouleurs galaxies are slightly rounder, on average,
than blue de Vaucouleurs galaxies. By contrast, blue galaxies
with exponential profiles appear very
much flatter, on average, than red galaxies with
exponential profiles. The red exponential galaxies are primarily
disk galaxies seen nearly edge-on, with reddening due
to the presence of dust, rather than to an intrinsically
red stellar population.

\end{abstract}


\section{Introduction}
Observationally based estimates of the three-dimensional shapes of
galaxies serve as a diagnostic of the physics of galaxy formation
and evolution. Galaxies can only be seen in projection against the
sky; thus, astronomers can only attempt to deduce their three-dimensional
properties from their two-dimensional, projected properties.
Obviously, information about intrinsic shapes is lost in projection;
for instance, using only the two-dimensional surface photometry of
a given galaxy, it is impossible to determine its intrinsic
three-dimensional shape.

Elliptical galaxies have isophotes that are well approximated
as ellipses (hence the name ``elliptical''). The shape of
an ellipse is specified by its axis ratio $q$, with
$0 \leq q \leq 1$. The three-dimensional isophotal
surfaces of elliptical galaxies are generally modeled as
ellipsoids. A stellar system whose isophotal surfaces
are similar, concentric ellipsoids, without axis twisting,
will have projected isophotes which are similar, concentric
ellipses, without axis twisting \citep{co56,st77}. The apparent
axis ratio $q$ of the projected ellipses depends on the
viewing angle and on the intrinsic axis ratios $\beta$ and
$\gamma$ of the ellipsoid. Here, $\beta$ is the ratio of
the intermediate to long axis, and $\gamma$ is the ratio
of the short to long axis; thus, $0 \leq \gamma \leq \beta \leq 1$.

Beginning with \citet{hu26}, many attempts have been made to
deduce the distribution of intrinsic shapes of elliptical galaxies,
given their distribution of apparent shapes. The early assumption,
in the absence of evidence to the contrary, was that elliptical
galaxies were oblate spheroids, flattened by rotation \citep{sa70}.
If elliptical galaxies were all oblate spheroids ($\beta = 1$),
or all prolate spheroids ($\beta = \gamma$), and if their orientations
were random, then it would be possible to deconvolve their
distribution of apparent axis ratios $f(q)$, to find their
distribution of intrinsic axis ratios $N(\gamma)$. However,
the pioneering work of \citet{be75} and \citet{il77} led astronomers
to abandon the assumption that elliptical galaxies are necessarily
oblate. The shapes of ellipticals have been reanalyzed with the
assumption that they are intrinsically prolate or triaxial, rather than
oblate \citep{bi78, be80, bi80, bi81, ry96}.

Statements about the intrinsic shapes of galaxies must be statistical
in nature, since astronomers do not exactly know the distribution
$f(q)$ of axis ratios for a given class of galaxy. In this paper,
we will be examining the apparent axis ratios for galaxies in
the Sloan Digital Sky Survey(hereafter SDSS; York {\em et al.} 2000). 
The set of axis ratios we analyze
constitute a finite sample drawn from a parent population $f(q)$.
In this paper, we take into account the finite size of the
sample in rejecting or accepting, at a known confidence level,
two null hypotheses: that the galaxies are randomly oriented
oblate spheroids, or that they are randomly oriented prolate
spheroids. To accomplish this, we make a kernel estimate,
${\hat f}(q)$, of the distribution of axis ratios and
mathematically invert ${\hat f}(q)$ to find ${\hat N_o}(\gamma)$
and ${\hat N_p}(\gamma)$, the estimated distribution of
intrinsic axis ratios for a population of oblate spheroids
and a population of prolate spheroids, respectively.

The rest of the paper is organized as follows. In section~\ref{data},
we describe the Sloan Digital Sky Survey, and the methods by which
the apparent axis ratios of the galaxies are estimated. In
section~\ref{method}, we present a brief review of the nonparametric
kernel estimators used in this paper. In sections~\ref{deV} and
\ref{exp}, we find the kernel estimate ${\hat f}(q)$ for galaxies
with de Vaucouleurs luminosity profiles and for galaxies with
exponential profiles, and find the implications for their
intrinsic shapes. In section~\ref{discuss}, we discuss our results.

\section{Data}
\label{data}
The Sloan Digital Sky Survey is
a digital photometric and spectroscopic survey which will, when
completed, cover one quarter of the celestial sphere in the North
Galactic hemisphere, and produce a smaller ($\sim 225 \,\sq\degr$)
but much deeper survey in the South Galactic hemisphere. The
photometric mosaic camera (\citep{gu98}
; see also Project Book \S 4\footnote{
http://www.astro.princeton.edu/PBOOK/welcome.htm}, 
`` The Photometric Camera'') images the sky by
scanning along great circles at the sidereal rate. The imaging
data are produced simultaneously in five photometric bands
($u'$, $g'$, $r'$, $i'$, and $z'$; \citet{fu96}) with effective wavelengths 
of 3543, 4770, 6231, 7625, and 9134 \AA\footnote{ We refer to the 
measured magnitudes in this paper as $u^*$, $g^*$, $r^*$, $i^*$, and $z^*$ 
because the absolute calibration of the SDSS photometric system is still 
uncertain at the $\sim 0.03^m$ level. The SDSS filters themselves are 
referred to as $u'$, $g'$, $r'$, $i'$, and $z'$.  All magnitudes are given on 
the $AB_{\nu}$ system \citep{ok83}. For additional discussion 
regarding the SDSS photometric system see \citet{fu96} and \citet{fa99}.}. 

In June 2001, the SDSS presented Early Data Release (EDR) to the 
general astronomical community, consisting 462 square degrees of 
imaging data in five bands. The data are acquired 
in three regions; along the celestial equator in the Southern galactic 
sky; along the celestial equator in the northern galactic sky; and in a 
region overlapping the Space Infrared Telescope Facility First Look Survey.
Galaxies in the EDR were analyzed with the
SDSS photometric pipeline {\it Photo}~\citep{lu01}. This code fits
two models to the two-dimensional image of each galaxy. One
model has a de Vaucouleurs profile:
\begin{equation}
I(r) = I_0 \exp \left( - 7.67[(r/r_e)^{1/4}] \right)
\label{eq:dev}
\end{equation}
which is truncated beyond $7 r_e$ to go smoothly to zero at
$8 r_e$, and with some softening within $r_e/50$. The second
model has an exponential profile:
\begin{equation}
I(r) = I_0 \exp ( - 1.68 r / r_e )
\label{eq:exp}
\end{equation}
which is truncated beyond $3 r_e$ to go smoothly to zero
at $4 r_e$. Each model is assumed to have concentric isophotes
with constant position angle $\phi$ and axis ratio $q$. Before
the model is fit to the data, the model is convolved with
a double-Gaussian fit to the point spread function (PSF). Assessing
each model with a $\chi^2$ fit gives $r_e$, $q$, and $\phi$ for the
best-fitting model, as well as $P({\rm deV})$ and $P({\rm exp})$, the
likelihood associated with the best-fitting de Vaucouleurs
and exponential model, respectively.

We use the model fits in the $r^*$ band to divide the galaxies
into two classes: the ``de Vaucouleurs'' galaxies are those
with $P({\rm deV}) > P({\rm exp})$ and the ``exponential''
galaxies are those with $P({\rm exp}) > P({\rm deV})$. In our data 
analysis we chose the likelihood $P>10^{-5}$ for well behaved distribution
of galaxies and the spectroscopic redshift $z<0.2$, to reduce the 
effects of gravitational lensing of foreground objects. In addition, 
we require that a fit using one of the galaxy models is better than a 
pure PSF fit. The spectroscopic sample of the EDR contains 13092 de 
Vaucouleurs galaxies, and 6081 exponential
galaxies, based on these criteria. Although 
the classification in the SDSS is based purely on the surface
brightness profile, it is generally true that galaxies classified
as ``elliptical'' in the standard morphological schemes are
better fitted by de Vaucouleurs profiles than by exponential
profiles~\citep{ko89}, while galaxies morphologically classified
as ``spiral'' are better fitted by exponential profiles.

The SDSS photometric analysis also provides an independent
measure of the axis ratio; one based on the second moments of
the surface brightness distribution. The Stokes
parameters $Q$ and $U$ are given in terms of the
flux-weighted second moments as
\begin{eqnarray}
Q & \equiv & \langle x^2/r^2 \rangle - \langle y^2/r^2 \rangle \\
U & \equiv & \langle xy/r^2 \rangle
\end{eqnarray}
If the isophotes of the galaxy are indeed concentric ellipses
of constant position angle and axis ratio,
then the axis ratio $q_{\rm Stokes}$ is
related to the values of $Q$ and $U$ by the relation
\begin{equation}
q_{\rm Stokes} = { \sqrt{Q^2 + U^2} - 1 \over \sqrt{Q^2 + U^2} + 1} \ .
\end{equation}
Unlike the axis ratios $q_{\rm model}$ found by fitting models, the
values of $q_{\rm Stokes}$ do not attempt to correct for the effects
of seeing.

\begin{figure}
\vspace{2.5in}
\includegraphics{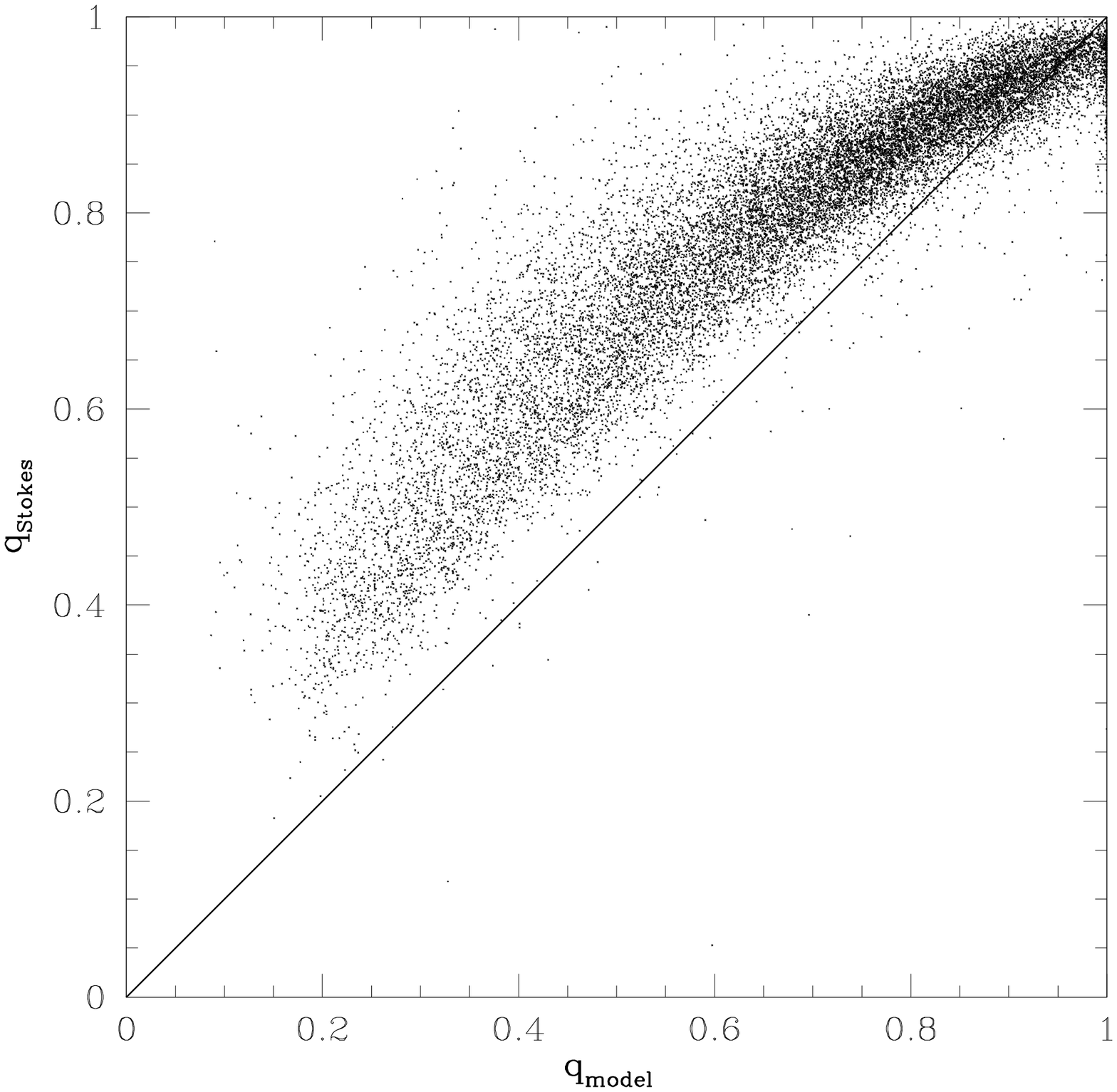}
\caption{
Axis ratio $q_{\rm Stokes}$ vs. axis ratio $q_{\rm model}$
for the galaxies studied.
}
\label{qsqm}
\end{figure}

\begin{figure}
\vspace{2.5in}
\includegraphics{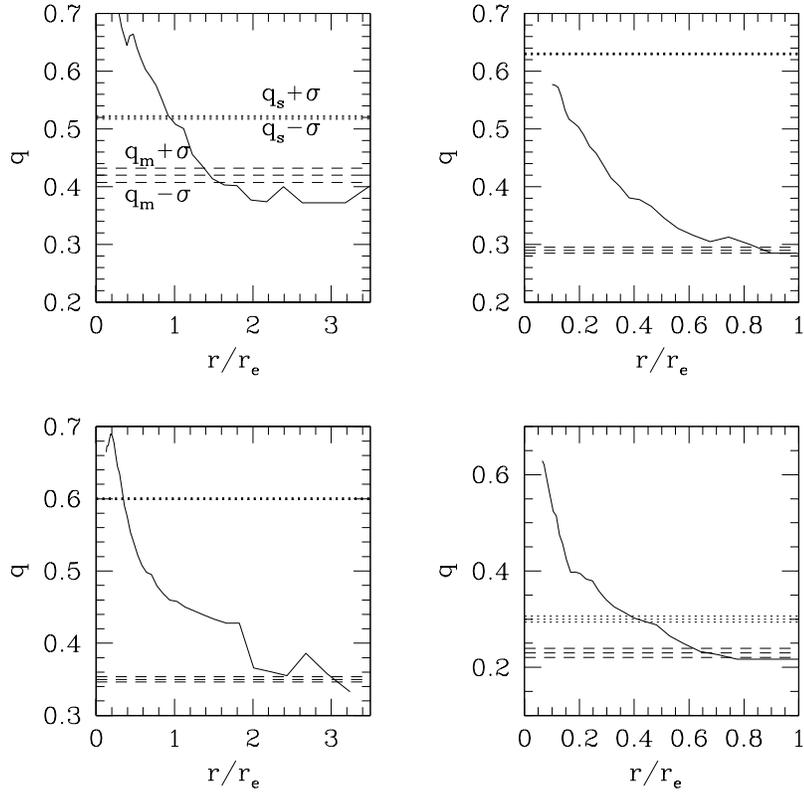}
\caption{
Axis ratio $q$ as a function of $r/r_e$, for a sample
of four de Vaucouleurs galaxies (two larger and two medium in size) 
in the SDSS. The solid line is the axis ratio found by 
the IRAF isophote fitting routine. The dotted lines represent
$q_{\rm Stokes}$ plus and minus the associated error $\sigma$.
The dashed lines represent $q_{\rm model}$ plus and minus
the associated error $\sigma$.
}
\label{qvsr}
\end{figure}

For the galaxies in the SDSS EDR, the value of
$q_{\rm Stokes}$ is generally larger than $q_{\rm model}$ as shown
in Figure~\ref{qsqm}. To investigate the origin of this difference, we used the
elliptical isophote fitting routine in Imaging and Reduction Analysis
Facility (IRAF) to plot $q$ versus $r/r_e$ for a subset of galaxies in
the sample. Figure~\ref{qvsr} shows the result for just
four of the de Vaucouleurs galaxies examined. The horizontal
dashed lines indicate $q_{\rm Stokes}$ plus and minus the
estimated error $\sigma$ in $q_{\rm Stokes}$;
the horizontal dotted lines indicate $q_{\rm model}$ plus
and minus the estimated error $\sigma$ in $q_{\rm model}$.
For these four galaxies, as for most galaxies in the SDSS,
$q_{\rm Stokes}$ is greater than $q_{\rm model}$.
As is also typical, $q(r)$
found by fitting individual isophotes decreases as a function of
radius; most elliptical galaxies are rounder in their central regions
than in their outer regions~\citep{ry01}.
The values of $q_{\rm Stokes}$,
based on the luminosity-weighted second moments, are primarily
indicating the axis ratio of the central regions of each galaxy,
where the surface brightness is highest. The values of $q_{\rm model}$,
by contrast, are more strongly influence by the axis ratio in
the outer regions. Since, in this paper, we are not primarily
interested in the central regions of the galaxies, we will adopt
$q_{\rm model}$ as our primary measure of the axis ratio.

The distribution of galaxies in the SDSS EDR is
strongly bimodal in the $g^* - r^*$ versus $u^* - g^*$ color-color
diagram \citep{st01}. The optimal color separation between
the two peaks is at $u^* - r^* = 2.22$. The detection of a
local minimum indicates that the two peaks correspond
roughly to early-type (E, S0) and late-type
(Sa, Sb, Sc, Irr) galaxies. The late-type galaxies
are the bluer group, reflecting their more recent
star formation activity. The color criterion of
\citet{st01} provides another means of dividing
our data set. In addition to distinguishing between
de Vaucouleurs and exponential galaxies, we can
also distinguish between red ($u^* - r^* > 2.22$)
and blue ($u^* - r^* \leq 2.22$) galaxies. Unsurprisingly,
the de Vaucouleurs galaxies are predominantly red
and the exponential galaxies are predominantly blue.
Of the 13092 de Vaucouleurs galaxies, 10898 
are red, but only 2194 are blue. Of the 6081
exponential galaxies, 4697 are blue, while only
1384 are red.

\section{Method}
\label{method}
We use a standard nonparametric kernel technique to estimate
the distribution of intrinsic axis ratios. General reviews
of kernel estimators are given by \citet{si86} and \citet{sc92};
applications to astronomical data are given by \citet{vi94},
\citet{tr95}, and \citet{ry96}. Here we give a brief overview,
adopting the notation of \citet{ry96}. Given a sample of
axis ratios for $N$ galaxies, $q_1$, $q_2$, $\dots$, $q_N$,
the kernel estimate of the frequency distribution $f(q)$ is
\begin{equation}
{\hat f} (q) = {1 \over N h} \sum_{i=1}^N K \left( { q - q_i \over
h } \right) \ ,
\end{equation}
where $K(x)$ is the kernel function, normalized so that
\begin{equation}
\int_{-\infty}^{+\infty} K(x) dx = 1 \ ,
\end{equation}
and $h$ is the kernel width, which determines the balance between
smoothing and noise in the estimated distribution. One way of
choosing $h$ is to use the value which minimizes the expected
value of the integrated mean square error between the true $f$
and the estimated $\hat {f}$ \citep{tr95}. We follow \citet{si86}
in using the formula
\begin{equation}
h = 0.9 A N^{-0.2} \ ,
\label{eq:silver}
\end{equation}
with $A = \min ( \sigma , Q_4 / 1.34 )$, where  $\sigma$ is the
standard deviation of the data and $Q_4$ is the interquartile
range. This formula for $h$ minimizes the expected value of
the mean square error for samples which are not strongly skewed
\citep{si86,vi94}. We choose a Gaussian kernel, to ensure that
$\hat f$ is smooth and differentiable.

To obtain physically reasonable results, with ${\hat f} = 0$
for $q < 0$ and $q > 1$, we apply reflective boundary
conditions \citep{si86, ry96}. In practice, this means
replacing the simple Gaussian kernel with the kernel
\begin{equation}
K_{\rm ref} = K \left( {q - q_i \over h} \right) +
K \left( { q + q_i \over h} \right) +
K \left( { 2 - q - q_i \over h} \right) \ .
\end{equation}
Use of this kernel assures the correct normalization,
$\int_0^1 {\hat f} (q) dq = 1$.

If we suppose that the $N$ galaxies in our sample
are randomly oriented oblate spheroids, then the
estimated frequency of intrinsic axis ratios, ${\hat N}_o (\gamma)$
can be found by the mathematical inversion
\begin{equation}
{\hat N}_o ( \gamma ) = { 2 \gamma \sqrt{1-\gamma^2} \over \pi}
\int_0^\gamma {d\ \over dq} \left( q^{-1} {\hat f}  \right)
{dq \over \sqrt{\gamma^2 - q^2} } \ .
\label{eq:obl}
\end{equation}
Similarly, if the galaxies are assumed to be randomly
oriented prolate spheroids, the estimated frequency
of intrinsic axis ratios, ${\hat N}_p (\gamma)$ is
given by
\begin{equation}
{\hat N}_p ( \gamma ) = { 2 \sqrt{1 -\gamma^2} \over \gamma \pi}
\int_0^\gamma {d\ \over dq} \left( q^2 {\hat f} \right)
{dq \over \sqrt{\gamma^2 - q^2}} \ .
\label{eq:pro}
\end{equation}
If the oblate hypothesis is incorrect, then the inversion
of equation~(\ref{eq:obl}) may result in ${\hat N}_o$ which is
negative for some values of $\gamma$. Similarly, if the prolate
hypothesis is incorrect, ${\hat N}_p$, from equation~(\ref{eq:pro}),
may be negative.

To exclude the oblate or prolate hypothesis at some statistical
confidence level, we must take into account the errors in $\hat f$
both from the finite sample size and from the errors in measuring
$q$ for individual galaxies. The error due to finite sampling
can be estimated by bootstrap resampling of the original data set.
Randomly taking $N$ data points, with replacement, from the
original data set, a new estimator $\hat f$ is created from
the bootstrapped data, and is then inverted to find new estimates
of ${\hat N}_o$ and ${\hat N}_p$. After creating a large number
of bootstrap estimates for $\hat f$, ${\hat N}_o$, and ${\hat N}_p$,
error intervals can be placed on the original kernel estimates.
In this paper, we did 300 bootstrap resamplings of each data set.
An additional source of error in $\hat f$, ${\hat N}_o$, and ${\hat N}_p$
is the error that is inevitably present in the measured values of the
apparent axis ratio. The SDSS EDR galaxies have an error $\sigma_i$
associated with the axis ratio $q_i$ of each galaxy. 
To model the
effect of errors, we replace the kernel width $h$ given by
equation~(\ref{eq:silver}) with a broader width
\begin{equation}
{h'}_i = \sqrt{h^2 + \sigma_i^2} \ .
\end{equation}

\section{Galaxies with de Vaucouleurs Profiles}
\label{deV}

\begin{figure}
\vspace{2.5in}
\includegraphics{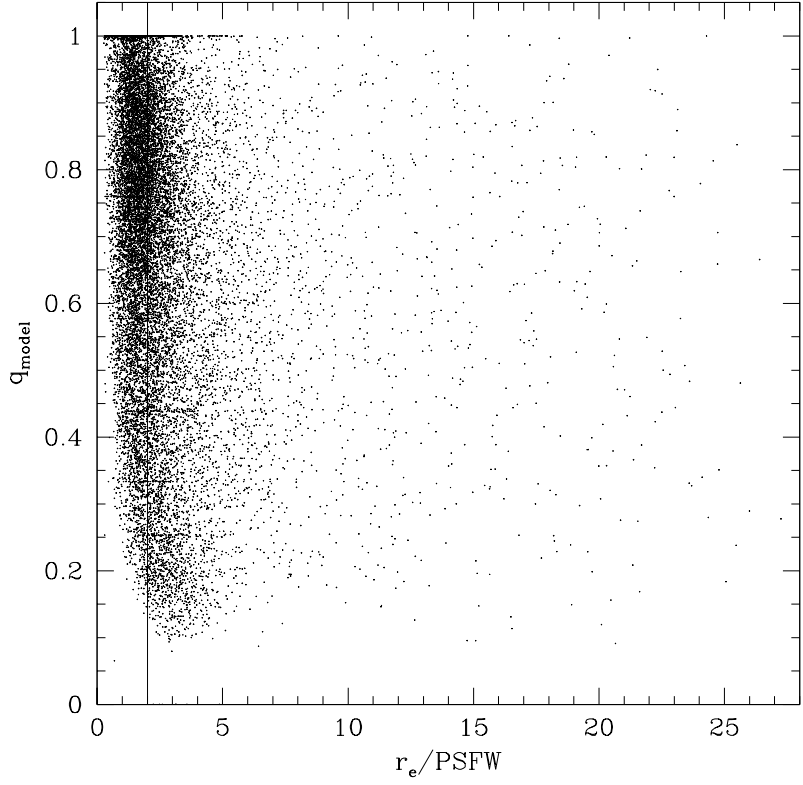}
\caption{
Axis ratio ratio q  as a function of effective radius,
measured in units of the PSFW, for galaxies
with both de Vaucouleurs and exponential profiles. The solid line 
is drawn at $r_e/PSFW = 2$. 
 }
\label{qmrepsfw}
\end{figure}

\begin{figure}
\vspace{2.5in}
\includegraphics{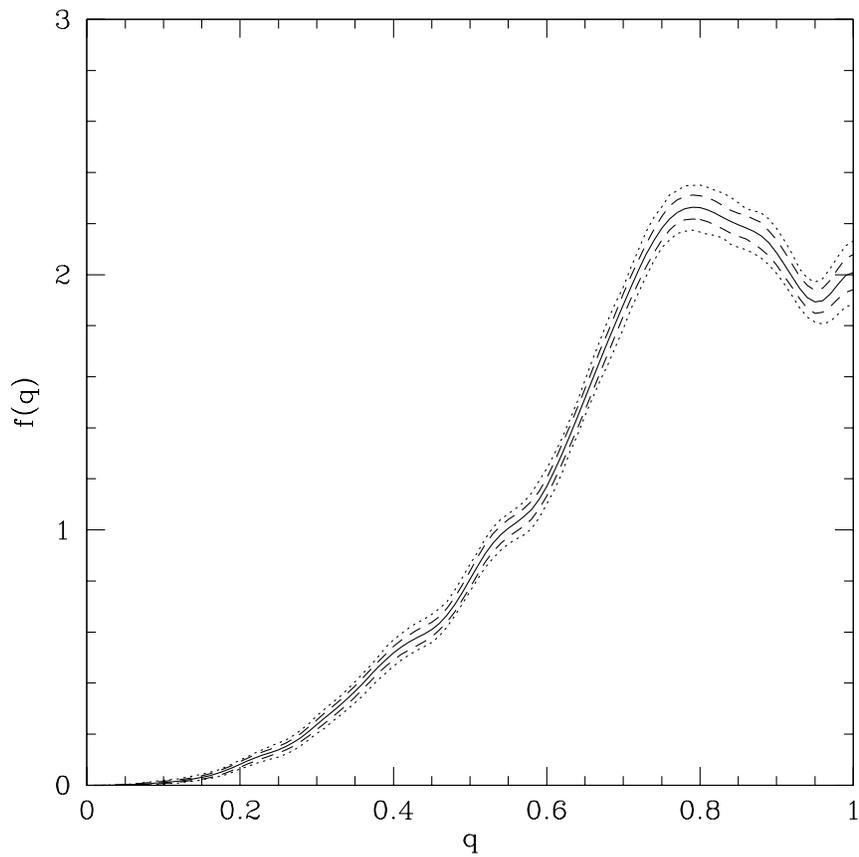}
\caption{Nonparametric kernel estimate of the distribution
of apparent axis ratios for the subsample of 10{,}898 red de Vaucouleurs
galaxies. The kernel width is $h = 0.026$.
}
\label{allred}
\end{figure}

We can now apply the mathematical apparatus described in the
previous section to our four subsamples of galaxies: red
de Vaucouleurs galaxies, blue de Vaucouleurs galaxies, red
exponential galaxies, and blue exponential galaxies. Unfortunately,
given the desirability of large $N$ in determining $f(q)$,
not all the galaxies in the SDSS EDR are sufficiently well-resolved
for their axis ratios to be reliably determined. A plot of $q_{\rm model}$ 
versus $r_e$, measured in units of
the PSF width (PSFW), as shown in Figure~\ref{qmrepsfw}, reveals that 
the galaxies with $q_{\rm model} = 1$ are
mainly galaxies whose effective radius is not much larger
than the PSF width (typically, $r_e \lesssim 2 {\rm\,PSFW}$). For instance,
Figure~\ref{allred} shows the estimated distribution $\hat{ f}(q_{\rm model})$ for all the red de Vaucouleurs galaxies ($N = 10{,}898$). 
In addition to the main peak at $q_{\rm model} \approx 0.80$,
there is a secondary peak at $q_{\rm model} = 1$. 
To eliminate these spuriously round, poorly resolved galaxies,
we retain in our samples only those galaxies with $r_e > 2 {\rm\,PSFW}$.
After this purge of too-small galaxies, there are
$N = 5659$ galaxies in the red de Vaucouleurs subsample,
$N = 1784$ galaxies in the blue de Vaucouleurs subsample,
$N = 815$ galaxies in the red exponential subsample,
and $N = 2263$ galaxies in the blue exponential subsample.

\begin{figure}
\vspace{2.5in}
\includegraphics{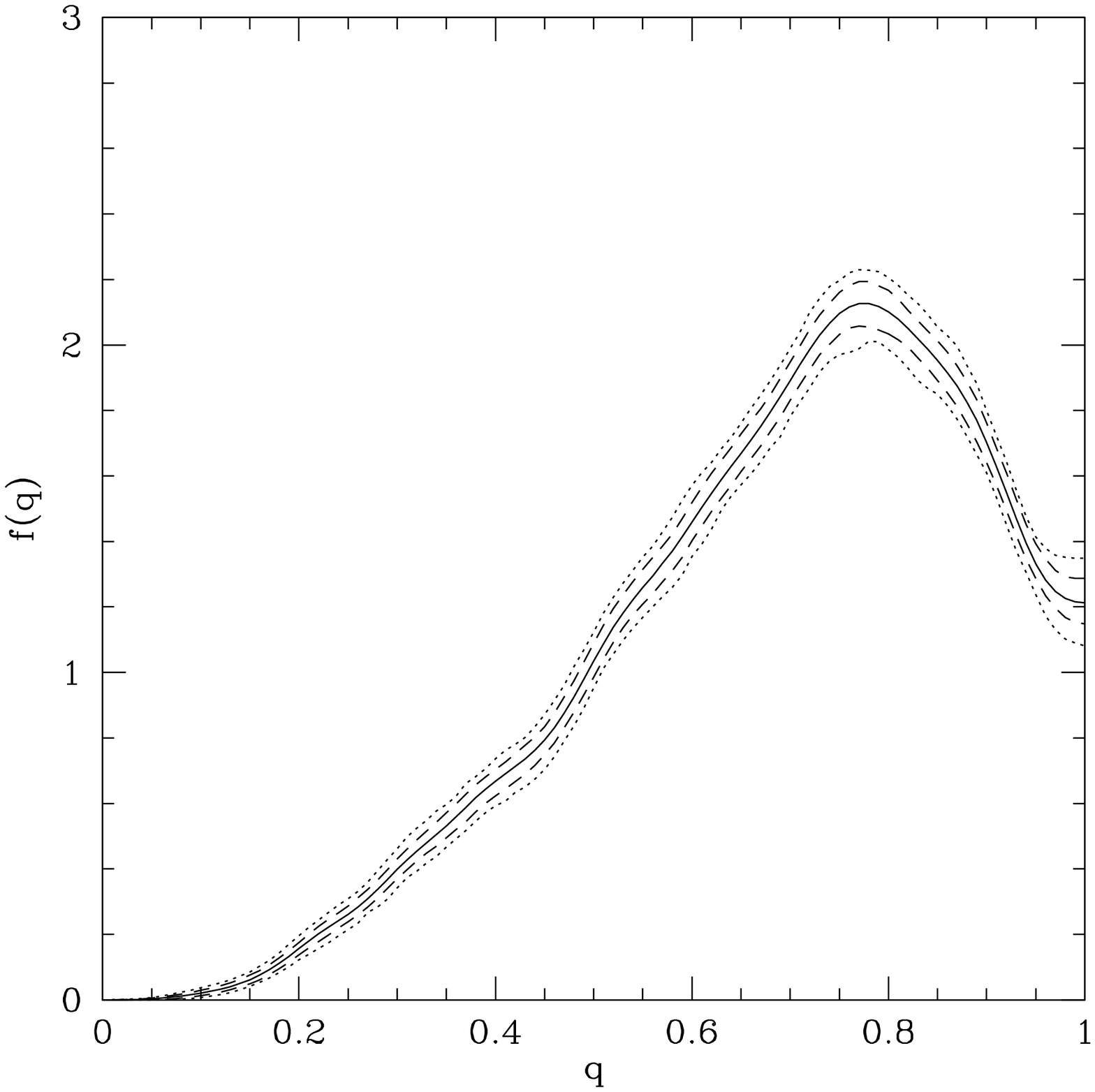}
\vspace{2.5in}
\includegraphics{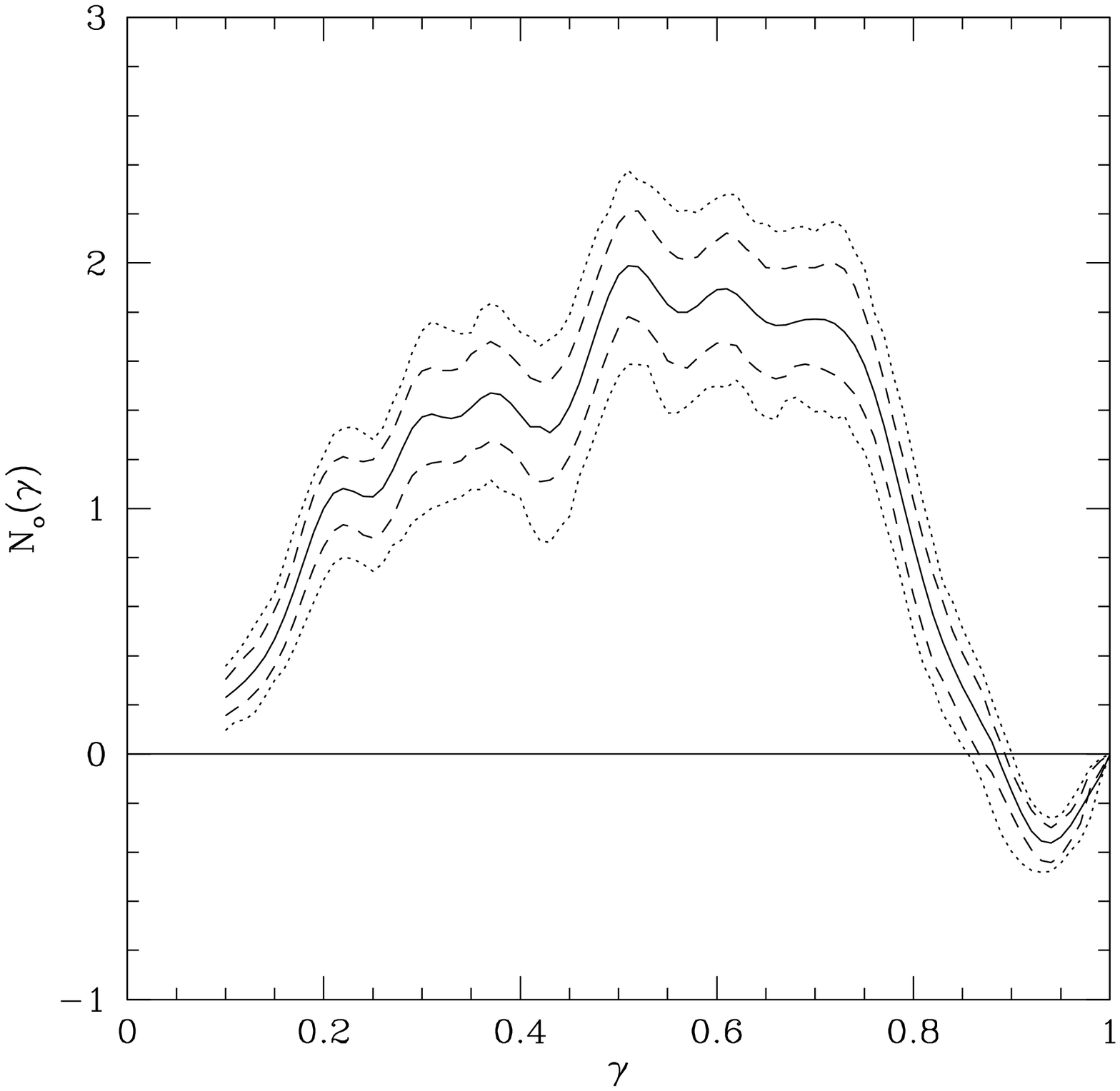}
\vspace{2.5in}
\includegraphics{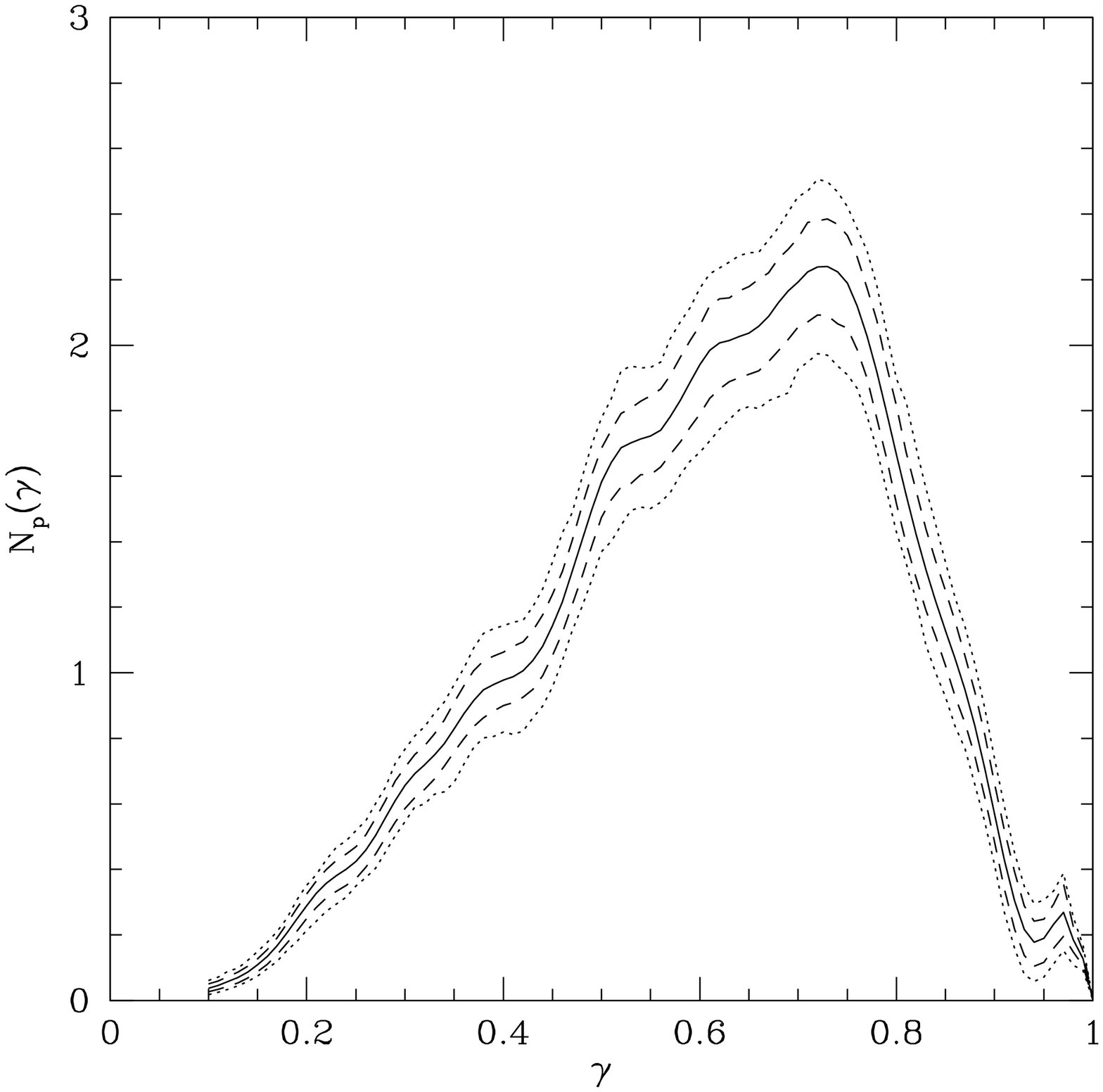}
\caption{Top: The distribution of apparent axis ratios for
a subsample of 5659 red de Vaucouleurs galaxies with
$r_e > 2 {\rm\,PSFW}$.
Middle: The distribution of intrinsic axis ratios for the
same subsample, assuming the galaxies are randomly oriented
oblate objects.
Bottom: The distribution of intrinsic axis ratios, assuming
the galaxies are randomly oriented prolate objects.
The kernel width is $h = 0.03$. The solid line in each
panel is the best fit, the dashed lines give the 80\%
error interval, and the dotted lines give the 98\%
error interval.
}
\label{reddev}
\end{figure}

The estimated distribution ${\hat f} (q_{\rm model})$ for
the edited subsample of red de Vaucouleurs galaxies (excluding
galaxies with $r_e < 2 {\rm\,PSFW}$) is shown as the
solid line in the upper panel of Figure~\ref{reddev}.
Note that the secondary peak at $q_{\rm model} = 1$ has disappeared.
The dashed lines to either side of the solid line are the
80\% confidence intervals, estimated by bootstrap resampling,
while the dotted lines are the 98\% confidence interval.
The estimated distribution ${\hat N}_o (\gamma)$ of intrinsic
axis ratios, given the oblate hypothesis, is shown in the
middle panel of Figure~\ref{reddev}. The 98\% confidence
interval drops below zero for $\gamma > 0.91$; thus the
oblate hypothesis for this subsample of galaxies can be
rejected at the 99\% (one-sided) confidence interval.
To produce as few nearly circular galaxies as are seen,
there would have to be a negative number of nearly
spherical oblate galaxies.
The estimated distribution ${\hat N}_p (\gamma)$, given
the prolate hypothesis, is shown in the bottom panel
of Figure~\ref{reddev}. The best estimate for ${\hat N}_p$,
shown as the solid line, is positive everywhere. 
Thus, the surface photometry of the red de Vaucouleurs
galaxies is consistent with their being a population
of randomly oriented prolate spheroids. If they are all
prolate, their average intrinsic axis ratio is
\begin{equation}
\langle \gamma \rangle_p = \int_0^1 \gamma {\hat N}_p (\gamma) d\gamma 
= 0.608 \ .
\end{equation}
Although the shape distribution for red de Vaucouleurs galaxies is
consistent with the prolate hypothesis, it doesn't require
prolateness. The galaxies could also be triaxial and produce
the same distribution of apparent shapes.

\begin{figure}
\vspace{2.5in}
\includegraphics{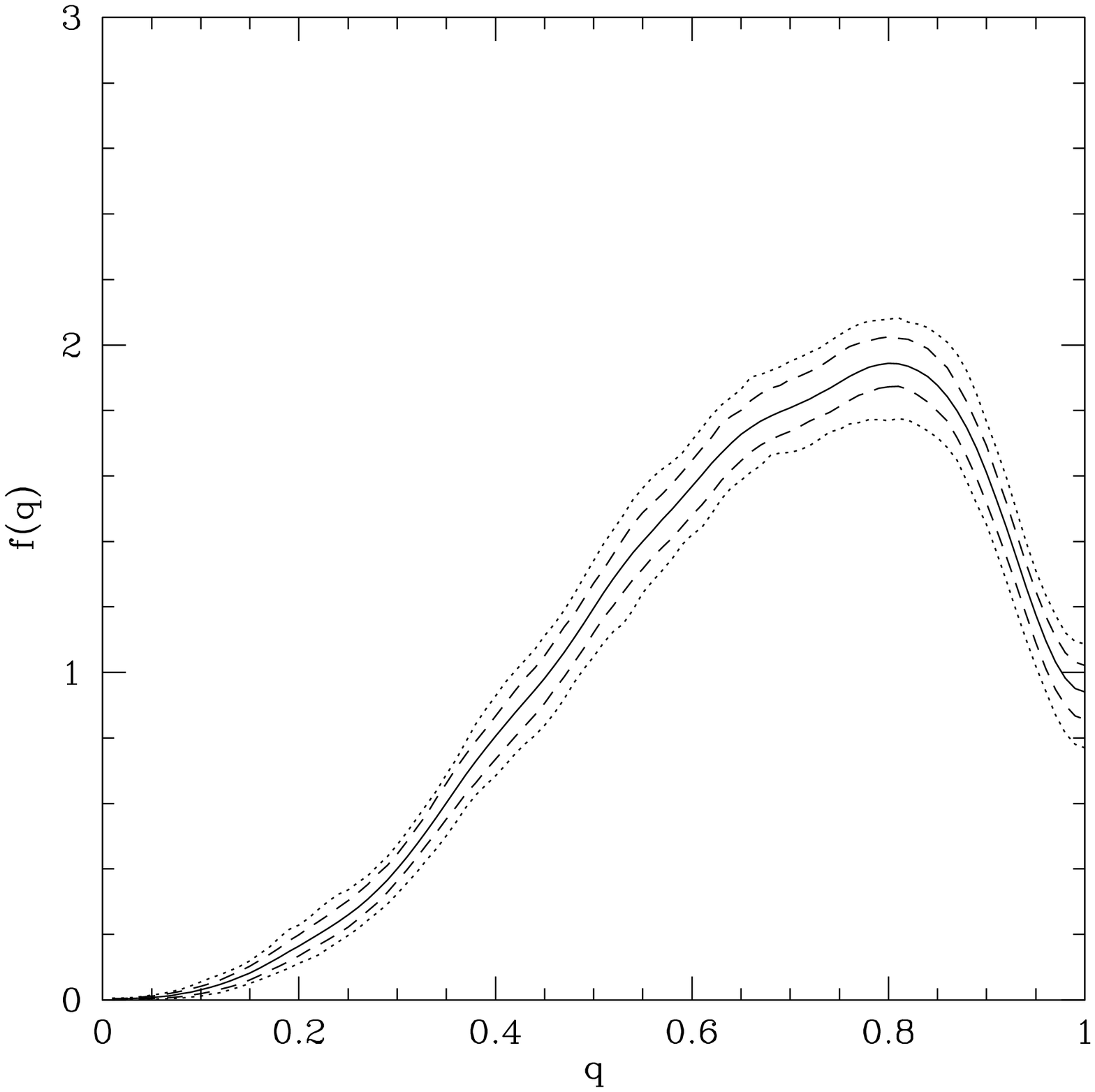}
\vspace{2.5in}
\includegraphics{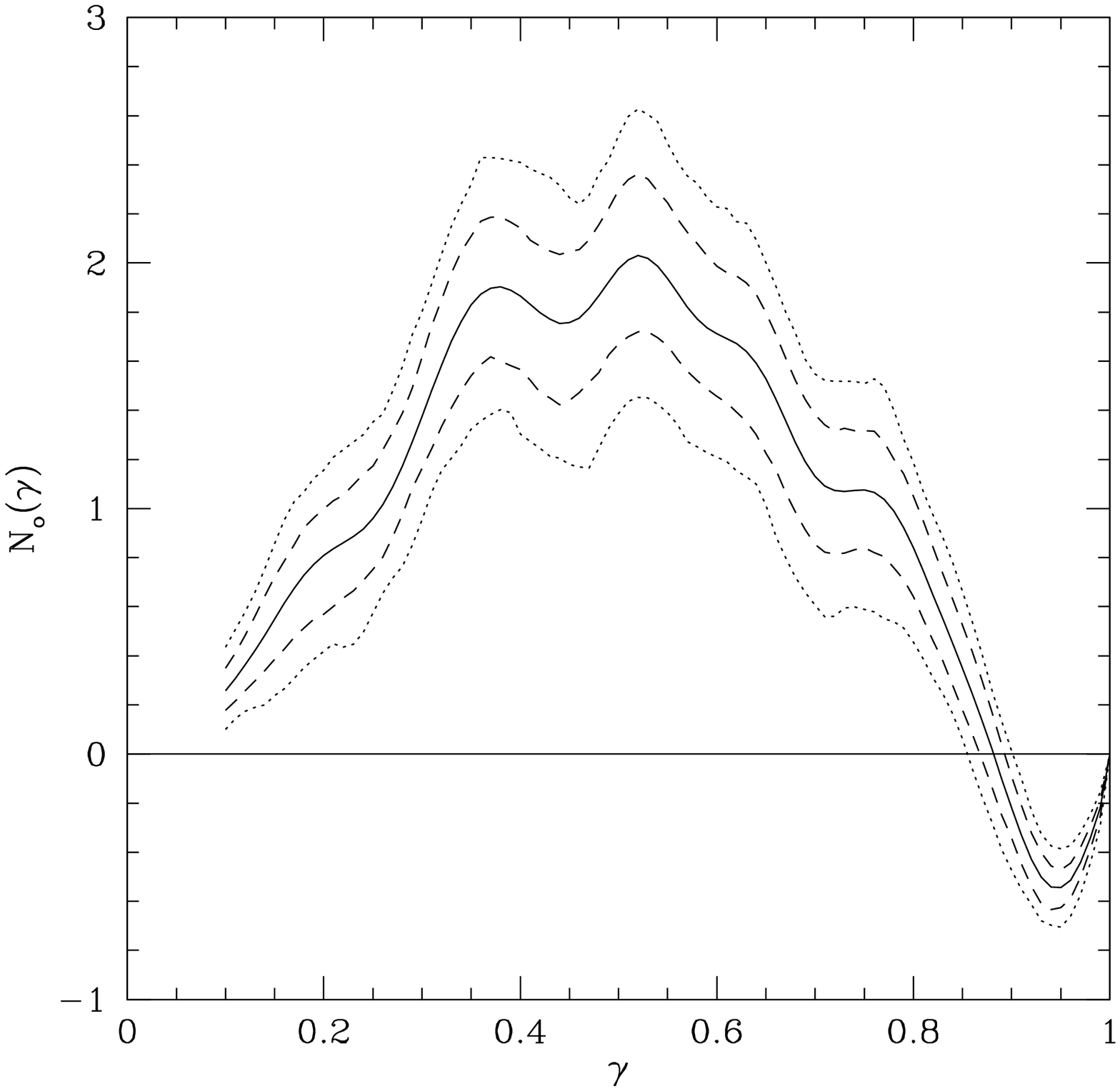}
\vspace{2.5in}
\includegraphics{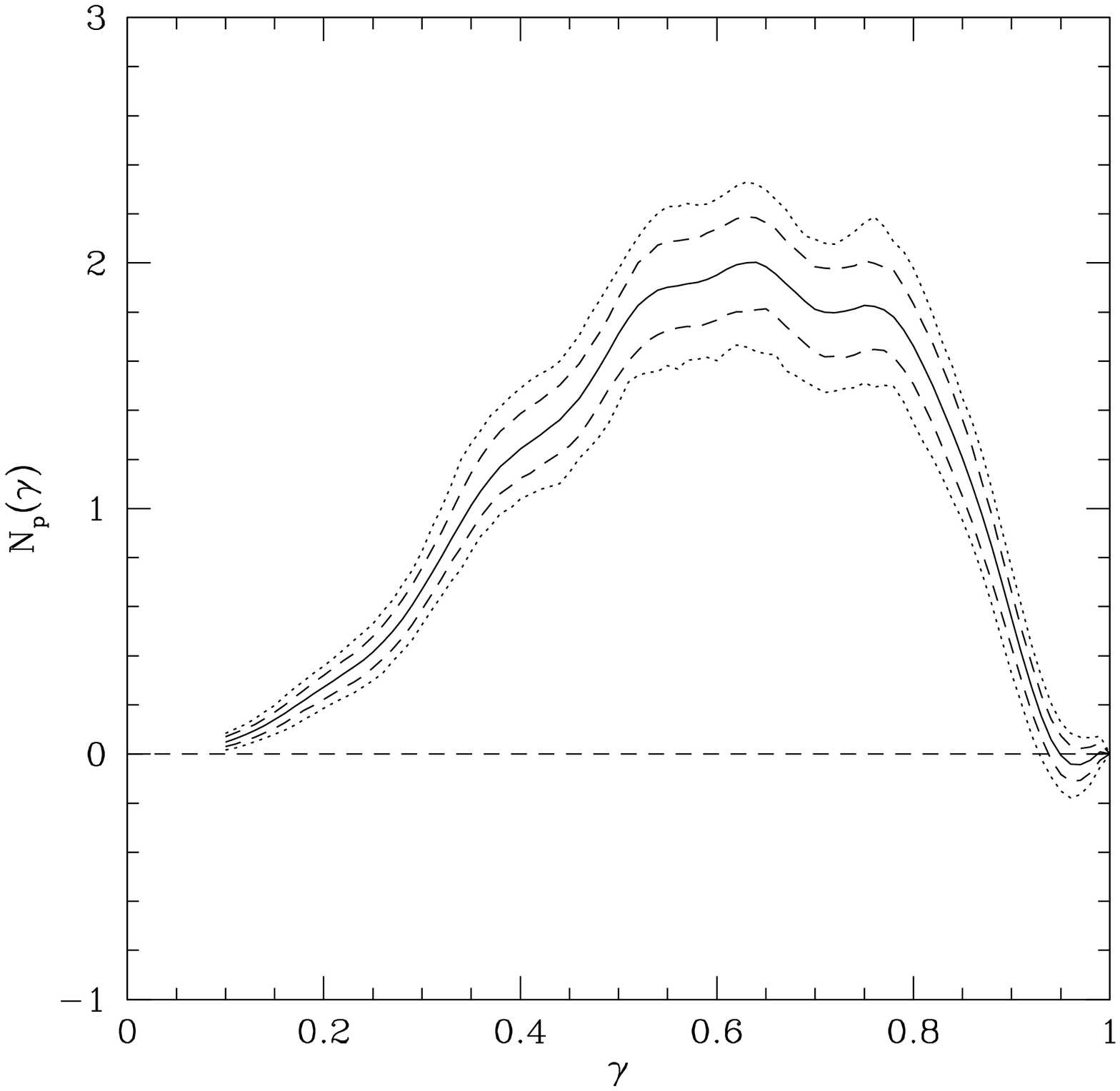}
\caption{As in Figure~\ref{reddev}, but for a subsample of 1784 blue
galaxies well fit by a de Vaucouleurs profile. The kernel
width is $h = 0.042$.
}
\label{bluedev}
\end{figure}

Plots of $\hat f$, ${\hat N}_o$, and ${\hat N}_p$ are
given in Figure~\ref{bluedev} for the 1784 blue de Vaucouleurs
galaxies. Just as for the red de Vaucouleurs galaxies,
the oblate hypothesis can be rejected at the 99\% confidence
level, due to the scarcity of nearly circular galaxies in projection.
The prolate hypothesis, though, cannot be rejected 
at the 90\% confidence level for this sample; see the bottom panel of 
Figure~\ref{bluedev}. If the blue de Vaucouleurs galaxies are prolate, 
then ${\hat N}_p$ yields an average intrinsic axis ratio $\langle \gamma \rangle_p = 0.592$, only slightly smaller than that for red de Vaucouleurs 
galaxies. As a measured by  Kolmogorov-Smirnov (K-S) test, however, 
the distribution of $q$ for blue de Vaucouleurs galaxies is  significantly 
different from that for red de Vaucouleurs. The K-S probability from
comparing the two samples is $P_{KS}= 3 \times 10^{-3}$;
i.e., the two samples are different at the
99.7\% confidence level. 

\section{Galaxies with Exponential Profiles}
\label{exp}

Unlike galaxies with de Vaucouleurs profiles, which are generally
smooth ellipticals, galaxies with exponential profiles are generally
spiral galaxies, containing nonaxisymmetric structures such as spiral
arms and bars. Given such a large amount of substructure present
in spiral galaxies, attempting to characterize their shape by a
single axis ratio $q$ is a gross oversimplification. Nevertheless,
as long as it is not over-analyzed, the distribution ${\hat f}(q)$
contains useful information about the overall shape of exponential
galaxies.

\begin{figure}
\vspace{2.5in}
\includegraphics{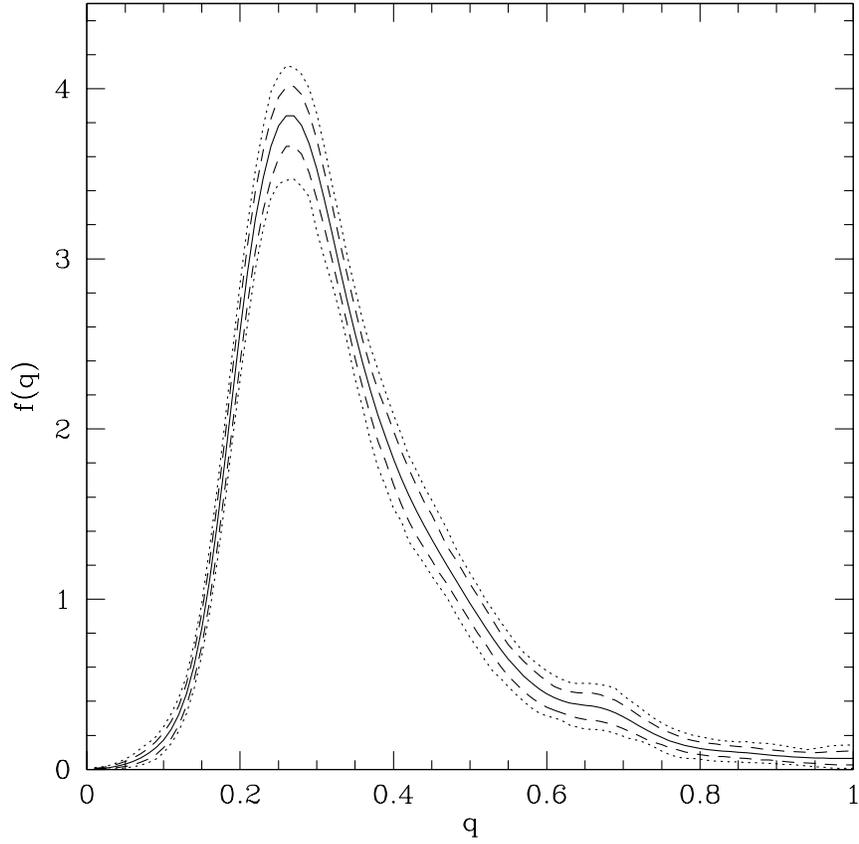}
\caption{The distribution of apparent axis ratios for a
subsample of 815 red galaxies well fit by an exponential
profile (only galaxies with $r_s > 2 PSFW$ are included). The kernel width is 
$h = 0.040$.
}
\label{redexp}
\end{figure}

For instance, Figure~\ref{redexp} shows the estimated value of
${\hat f} (q_{\rm model})$ for the 815 red exponential galaxies.
The peak in $\hat f$ is at $q_{\rm model} \approx 0.27$, and
relatively few red exponential galaxies have $q_{\rm model} > 0.6$.
The large apparent flattening of the galaxies in the red exponential
subsample is a sign that they are not a population of randomly oriented
disks. Instead, the sample preferentially contains edge-on, or nearly
edge-on disks. The red color of galaxies in this subsample is not
the intrinsic color of the stars, but rather is the result of internal
reddening by dust in the galaxy's disk. For  a typical 
edge-on spiral galaxy like NGC 4594, the maximum reddening $E(B-V)$ 
is 0.4   \citep{kn91}. Then the corresponding reddening 
in the $u-r$ color will  be $\sim$ 1, using the transformation $E(u-r)
= [A_u/E(B-V) - A_r/E(B-V)]E(B-V)$ with $A_u/E(B-V)= 5.155$ and 
$A_r/E(B-V)= 2.751$ \citep{s01}.   

\begin{figure}
\vspace{2.5in}
\includegraphics{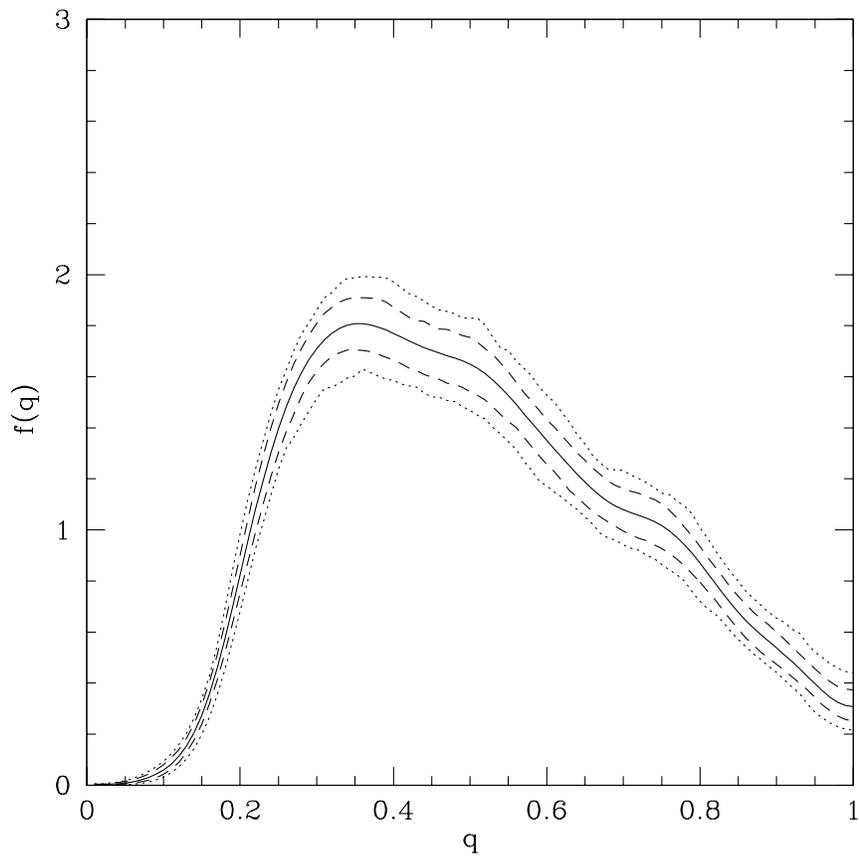}
\caption{As in Figure~\ref{redexp}, but for a subsample of 2263 blue
galaxies well fit by an exponential profile. The kernel width is $h = 0.038$.
}
\label{blueexp}
\end{figure}

By contrast, Figure~\ref{blueexp} shows the estimated value of
${\hat f} (q_{\rm model})$ for the 2263 blue exponential galaxies.
The apparent shapes of blue exponential galaxies are very different
from the apparent shapes of red exponential galaxies; a K-S
test comparing the two populations yields $P_{\rm KS} = 5.10^{-95}$. 
The scarcity of exponential galaxies with $q \gtrsim 0.9$ is an
indication that the exponential galaxies are not perfectly axisymmetric
disks. Indeed, visual inspection of the SDSS images reveals
that most of the exponential galaxies contain readily visible
nonaxisymmetric structure, in the form of spiral arms, bars,
or tidal distortions. 

\section{Discussion}
\label{discuss}
The galaxies in the SDSS EDR with de Vaucouleurs profiles have
a distribution of apparent shapes which is incompatible (at the
99\% confidence level) with their being randomly oriented oblate
spheroids. This is consistent with the result found by
\citet{la92} for a sample of 2135 elliptical galaxies with
shapes estimated from survey plates of the APM Bright Galaxy
Survey. When the SDSS survey is complete, it will provide
a sample of galaxies $\sim 20$ times larger than the
SDSS EDR. This increase in sample size will enable
us to determine more accurately the distribution
of apparent axis ratios $f(q)$. The kernel width
$h$ will be decreased by a factor $\sim 20^{-0.2}
\sim 0.55$. The error intervals, which are essentially
determined by the $N^{1/2}$ fluctuations in bins of
width $h$, will be reduced by a factor $\sim 20^{-0.4}
\sim 0.30$. Although a simple increase in the
sample size will not enable us to determine the
true distribution of intrinsic shapes, it will
enable us to make stronger statistical statements
about our rejection or acceptance of the
prolate or oblate hypothesis.

The blue de Vaucouleurs galaxies, with a mean axis
ratio of $\langle q_{\rm model} \rangle = 0.639$, are
only slightly flatter in shape than the red de Vaucouleurs
galaxies, with $\langle q_{\rm model} \rangle = 0.652$.
Thus, if the color of the blue de Vaucouleurs galaxies
is the result, at least in part, of recent star formation,
we can conclude that the overall shape of the galaxies
is not strongly affected by star formation.
Although elliptical galaxies with old stellar populations
tend to be rounder than those with young stellar populations
\citep{ry01}, this difference is only large at small radii
($r \lesssim r_e/8$), while the values of $q_{\rm model}$
used in this paper emphasize the axis ratio at much larger
radii ($r \gtrsim r_e$).

The galaxies with exponential profiles have, by contrast, shapes
which are strongly dependent on color, with the red exponential
galaxies consisting predominantly of dust-reddened edge-on (or
nearly edge-on) disks. A significant number of galaxies in the
SDSS EDR appear to be edge-on late-type galaxies, with exponential
profiles, rather than early-type galaxies, with de Vaucouleurs profiles.
The number of red exponential galaxies ($N = 815$) is 14.4\% of
the number of red de Vaucouleurs galaxies ($N = 5{,}659$). Thus,
if we attempted to select out elliptical galaxies purely on the
basis of color, we would have been faced with a significant contamination
by reddened disks. Spectroscopy or accurate surface photometry
are required to distinguish between elliptical galaxies and
disk galaxies. This analysis agrees very well with the result of 
\citet{sc99} and carries substantially more statistical weight.

In summary, galaxies with de Vaucouleurs profile have an axis
ratio distribution consistent, at a high confidence level, with
their being randomly oriented prolate spheroids (though it is
also consistent with their being triaxial systems). Galaxies
with exponential profiles have an axis ratio distribution which
is dependent on color, suggesting that red exponential galaxies
are nearly edge-on systems reddened by dust. Since a fair number of
red galaxies in the SDSS EDR are nearly edge-on exponential disks, it
is dangerous to select elliptical galaxies purely on the basis
of color.

\acknowledgments

We wish to thank Ani Thakar for his assistance in downloading
the data, and Brian Yanny, Jordi Miralda-Escud\'e, Richard Pogge and  
David Weinberg for their valuable suggestions. The Sloan Digital 
Sky Survey (SDSS) is a joint project of
The University of Chicago, the Institute of Advanced Study,
the Japan Participation Group, the Max-Planck-Institute for
Astronomy (MPIA), the Max-Planck-Institute for Astrophysics (MPA),
New Mexico State University, Princeton Observatory, the
United States Naval Observatory, and the University of Washington.
Apache Point Observatory, site of the SDSS telescopes, is operated
by the Astrophysical Research Consortium (ARC). Funding for the
project has been provided by the Alfred P. Sloan Foundation,
the SDSS member institutions, the National Aeronautics and
Space Administration, the National Science Foundation, the
U.S. Department of Energy, the Japanese Monbukagakusho, and
the Max Planck Society. The SDSS website is http://www.sdss.org/.

\clearpage

\clearpage



\clearpage

\clearpage

\clearpage

\clearpage

\clearpage

\clearpage

\end{document}